\documentclass[11pt,a4paper]{article}
\usepackage{jcappub}
\usepackage{amsmath,amssymb,graphics,epsfig,subfigure}
\usepackage{color}

\title{Strong field limit analysis of gravitational lensing in Kerr-Taub-NUT spacetime}

\author{Shao-Wen Wei,}
\author[1]{Yu-Xiao Liu,\note{Corresponding author}}
\author{Chun-E Fu,}
\author{Ke Yang}
\affiliation{Institution,\\Theoretical Physics, Lanzhou University, Lanzhou 730000, People's Republic of China}

\emailAdd{weishw@lzu.edu.cn, liuyx@lzu.edu.cn, fuche08@lzu.edu.cn, yangke09@lzu.edu.cn}

\abstract{
In this paper, we investigate the strong gravitational lensing by the stationary, axially-symmetric black hole in Kerr-Taub-NUT spacetime in the strong field limit. The deflection angle of light ray and other strong deflection limit coefficients are obtained numerically and they are found to be closely dependent on the NUT charge $n$ and spin $a$. The magnification and the positions of the relativistic images are computed. The caustics are studied and the results show that these caustics drift away from the optical axis, which is quite different from the Schwarzschild black hole case. Moreover, the intersections of the critical curves on the equatorial plane are obtained and it is shown that they increase with the NUT charge. These results show that there is a significant effect of the NUT charge on the strong gravitational lensing.}


\arxivnumber{1104.0776}

\begin{document}


\maketitle

\section{Introduction}

Deflection of light ray in a gravitational field is one of the consequences of Einstein's general theory of relativity. The phenomenon is referred to as gravitational lensing and the object causing a detectable deflection is called a gravitational lens. Researches on the gravitational lensing show that the value of the deflection angle of light depends on the observer-lens distance and the nature of the lens. This result implies that the gravitational lensing can provide us with the information about the distant stars. It also can help us to test the exotic objects in the universe and to estimate the values of the cosmological parameters \cite{Sutherland}. Moreover, it could provide a profound verification of alternative theories of gravity in the strong field regimes \cite{Bekenstein,Eiroa,Sarkar,Chen1}, and an effective detection of gravitational waves \cite{Stefanov,Wei} and the cosmic censorship hypothesis \cite{Virbhadra2,Virbhadra77}.

Gravitational lensing in weak field approximation has been developed in \cite{Schneider,Zakharov,Blandford}. In the weak field approximation, one can study the properties of the ordinary stars and galaxies. However, when a compact object (like a black hole) plays the role of the lens, there will be notable phenomenon very near the object, which can be described by the strong field limit or the strong deflection limit rather than the weak field approximation. The studies of the strong field limit lensing due to different black holes have received considerable attention in recent years, which suggests that we can extract the information of the black hole from lensing. Studies on it can be traced back to the work \cite{Darwin,Darwin2} of Darwin, as well as to that of Frittelli, Kling and Newman
\cite{Frittelli} and that of Virbhadra and Ellis \cite{Virbhadra}, where the authors presented a definition of an exact lens equation without reference to black hole background spacetime, and then the exact lens equation was constructed for the Schwarzschild black hole spacetime. It is worth noting that the Virbhadra-Ellis lens equation is effective for the case that the observer and the light source must be far away from the lens so that the gravitational fields there can be described by flat metric. Later, the work was extended to the Reissner-Nordstrom black hole lensing \cite{EiroaRomero} and the naked singularities lensing \cite{Virbhadra2,Virbhadra77}, which showed that the strong gravitational lensing by naked singularity is very different from that of black hole. With the lightlike geodesic equation, lensing in a spherically symmetric and static spacetime was considered in \cite{Perlick}.

Based on the Virbhadra-Ellis lens equation, Bozza \emph{et al}., \cite{BozzaCapozziello} proposed an analytical method for obtaining the deflection angle in the strong gravitational field case and the result shows that the deflection angle diverges logarithmically as the light rays get close to the photon sphere of a Schwarzschild black hole. In \cite{Bozza}, Bozza proved that the result is also held for other static spherically symmetric black hole lensing. He still extended the methods to study the spinning black holes \cite{Bozza67,Bozza72,Bozza74}. Moreover, Eiroa \emph{et al.}, have studied the gravitational lensing by the Reissner-Nordstrom black hole and the braneworld black hole \cite{Eiroa66,Eiroa71,Whisker}. Strong gravitational lensing by other black holes and wormholes were extensively investigated in \cite{Iyer,Hansen,Bhadra,Vazquez,Eiroa69,Nandi,Gyulchev75,Bozza76,Virbhadra79,Bisnovatyi,Yazadjiev,
Nun,Liu,Ghosh,Chengc28,Chengc29,Cheng27,Eiroa1011,Ding,Chen1102,Perlick2}.

On the other hand, the Kerr-Taub-NUT (KTN) black hole \cite{Demianski,Miller} is a remarkable solution of Einstein-Maxwell equations for electro-vacuum spacetime possessing with gravitomagnetic monopole. The KTN black hole carries three parameters, the mass $M$, the spinning parameter $a$ and the NUT charge $n$ known as the ``gravitomagnetic mass". The presence of the NUT charge gives rise to a black hole solution with some fascinating properties. For example, there exists no curvature singularity but conical singularities on its axis of symmetry, which is caused by the gravitomagnetic analogue of Dirac's string quantization condition \cite{Misner}. Furthermore, the influences of the NUT charge or gravitomagnetic mass $n$ on the spacetime structure, thermodynamics and geodesic equations, as well as the particles collision were studied in \cite{Bini,BiniCherubini,Aliev,AlievCebeci,Kagramanova,Abdujabbarov,
Morozova,LiuChen,Lammerzahl}.

The gravitomagnetic lensing in nonrotating NUT spacetime has been studied in \cite{NouriZonos}. The results show that the NUT charge influences the gravitational lensing through the null geodesics. The possibility to detect the gravitomagnetic masses with the next generation of microlensing experiments was analyzed in \cite{Morozova,Rahvar,KagramanovaKunz}. These results suggest that the influence of the NUT parameter will be stronger in the vicinity of compact gravitating objects with small radius. So the purpose of this paper is to study the effects of the NUT charge $n$ and spin $a$ on the strong gravitational lensing by the rotating supermassive KTN black hole.

The paper is structured as follows. The second section is devoted to the derivation of the first order differential system for the geodesics in the background of the KTN black hole. In Sec. \ref{SLKTN}, we study the lensing equation in the equatorial plane and calculate numerically the strong field limit coefficients and the deflection angle. And the differences of these coefficients between the Kerr black hole and the KTN black hole are analyzed. In Sec. \ref{QLKTN}, we study the quasi-equatorial lensing by the KTN black hole, where the precession of the orbit for small declinations, and the magnification of the images are obtained. The critical curves and caustic structure are analyzed in Sec. \ref{Critical}. And a brief discussion is given in Sec. \ref{Summary}.

\section{Geodesics in Kerr-Taub-NUT spacetime}
\label{CMenergy1}

The general KTN black hole solution of the Einstein field equations is described by the metric \cite{Demianski,Miller}
\begin{eqnarray}
ds^{2} &=& -\frac{\Delta}{\rho^{2}}\bigg[dt+(2n\cos\theta-a\sin^{2}\theta)d\phi\bigg]^{2} \nonumber \\
       &&+\frac{\sin^{2}\theta}{\rho^{2}}\bigg[adt-(r^{2}+n^{2}+a^{2})d\phi\bigg]^{2}
       +\frac{\rho^{2}}{\Delta}dr^{2}+\rho^{2}d\theta^{2},~~~~ \label{KTNmetric}
\end{eqnarray}
where
\begin{eqnarray}
 \rho^{2}&=&r^{2}+(n+a\cos\theta)^{2},\\
 \Delta&=&r^{2}-2Mr+a^{2}-n^{2}.
\end{eqnarray}
Here $M$ is the gravitoelectric mass of the black hole and $n$ is the NUT charge (or gravitomagnetic mass). The angular coordinates $\theta\in(0, \pi)$ and $\phi\in(0, 2\pi)$. In the particular case $n=0$, the solution (\ref{KTNmetric}) coincides with the Kerr black hole solution. And in the case $a=0$ and $n=0$, it recovers the Schwarzschild black hole solution.

The event horizon of the KTN black hole is located at the biggest root of the equation $\Delta=0$, i.e.,
\begin{eqnarray}
 r_{\textrm{h}}=M+\sqrt{M^{2}+n^{2}-a^{2}}.
\end{eqnarray}
For the case $a^2> {M^{2}+n^{2}}$, there will be no event horizon and a naked singularity will appear, which spoils the causality of the spacetime and is forbidden according to the Penrose's cosmic censorship conjecture. So, we here only consider the case $a^2\leq {M^{2}+n^{2}}$. Another important surface of the black hole is the ergosphere, which corresponds to $g_{tt}=0$. The outer one is
\begin{eqnarray}
 r_{\textrm{es}}=M+\sqrt{M^2+n^2-a^2 \cos^2\theta}.
\end{eqnarray}
Note that we always have $r_{\textrm{es}}\geq r_{\textrm{h}}$ and the inequality is saturated at $\theta=0,\;\pi$.

The geodesic equations can be solved with the Hamilton-Jacobi method for the metric (\ref{KTNmetric}) and the first-order differential system for the geodesic is given by (the motion of a charged particle around the KTN black hole immersed in an external magnetic field can be found in \cite{Abdujabbarov})
\begin{eqnarray}
 \rho^{2}\dot{r}=\pm\sqrt{R},\label{rdot}\\
 \rho^{2}\dot{\theta}=\pm\sqrt{\Theta},
\end{eqnarray}
\begin{eqnarray}
 \rho^{2}\dot{t}=&-&E\bigg[\sin^{2}\theta(a+2n\csc^{2}\theta)^{2}-4n(n+a(1+\cos\theta))\bigg]
                   -(2n \cos\theta \csc^{2}\theta-a)L\nonumber\\
                 &+&\frac{(r^{2}+a^{2}+n^{2})[E(r^{2}+a^{2}+n^{2})-aL]}{\Delta},\\
 \rho^{2}\dot{\phi}=&-&(a-2n\cot\theta\csc\theta)E+L\csc^{2}\theta+\frac{a[E(r^{2}+a^{2}+n^{2})-aL]}{\Delta},
  \label{Phidot}
\end{eqnarray}
where the dot denotes the derivative with respect to the affine parameter and $R$, $\Theta$ are
\begin{eqnarray}
 R&=&(E(r^{2}+a^{2})-a L)^{2}+n^{2}E(2a^{2}E-2aL+E(n^{2}+2r^{2}))\nonumber\\
        &&-\big[(L-aE)^{2}+m^{2}(n^{2}+r^{2})+\mathcal{K}\big]\Delta,\\
 \Theta&=&\mathcal{K}-\cos^{2}\theta[a^{2}(m^{2}-E^{2})+L^{2}\csc^{2}\theta]\nonumber\\
        &&+2n\cos\theta[2aE^{2}-am^{2}+2EL\csc^{2}\theta]-4n^{2}E^{2}\cot^{2}\theta.
\end{eqnarray}
Equations (\ref{rdot})-(\ref{Phidot}) are the first-order geodesic equations for a massive particle $m^{2}=1$ and for a photon $m^{2}=0$, respectively. $\mathcal{K}$ is a separation constant of motion. The constants $E$ and $L$ are the conservation of energy and orbital angular momentum
per unit mass of the motion and they correspond to the Killing fields $\partial_{t}$ and $\partial_{\phi}$, respectively. For the case $n=0$, Eqs. (\ref{rdot})-(\ref{Phidot}) just describe the motion of a particle in the Kerr black hole spacetime.
 
Moreover, we can express the lightlike geodesics in the following form
\begin{eqnarray}
 \int^{r}\frac{dr}{\pm\sqrt{R}}=\int^{\theta}\frac{d\theta}{\pm\sqrt{\Theta}},
\end{eqnarray}
\begin{eqnarray}
 \Delta\phi&=&a\int^{r}\frac{[E(r^{2}+a^{2}+n^{2})-aL]}{\pm\Delta\sqrt{R}}dr
            +\int^{\theta}\frac{L\csc^{2}\theta-(a-2n\cot\theta\csc\theta)E}
                {\pm\sqrt{\Theta}}d\theta,\\
 \Delta t&=&a\int^{r}\frac{(r^{2}+a^{2}+n^{2})[E(r^{2}+a^{2}+n^{2})-aL]}
                              {\pm\Delta\sqrt{R}}dr \nonumber \\
            &&+\int^{\theta}
              \frac{(a-2n \cos\theta \csc^{2}\theta)L-E[\sin^{2}\theta(a+2n\csc^{2}\theta)^{2}
                -4n(n+a(1+\cos\theta))]}
                {\pm\sqrt{\Theta}}d\theta.\nonumber \\
\end{eqnarray}
Note that the NUT charge $n$ indeed influences on the null geodesics, which will be shown to have significant effect on the black hole lensing.

As we know, in the real world, the graviational field far away from a compact object or even a black hole is very weak and can be described by a flat metric. So the compact object only significantly influences the motion of particle in the neighborhood of the object. On the other hand, we take the small value of $n/2M$ for the quasi-equatorial approximation with $\theta\sim \frac{\pi}{2}$. Then the light ray trajectory can be regarded as a straight line at infinity both in equatorial plane and quasi-equatorial approximation. From this view, one can identify the approximate light ray with three parameters $\psi_{o}$, $u$ and $h$. The first one $\psi_{o}$ represents the inclined angle that the incoming light ray forms with the equatorial plane. The second one $u$ is an impact parameter of the projection of the light ray trajectory in the equatorial plane. And the last one $h$ describes the height between the point of the projection closer to the black hole and the trajectory. Then following \cite{Vazquez,Bozza67,Gyulchev75}, if the observer is located at $(r_{o},\;\vartheta_{o})$ in the Boyer-Lindquist system, one can define two celestial coordinates $\zeta_{1}$ and $\zeta_{2}$ for an image. The coordinate $\zeta_{1}$ denotes the observable distance of the image with respect to the symmetry axis in direction normal to the ray of sight and the coordinate $\zeta_{2}$ measures the observable distance from the image to the source projection in the equatorial plane in the direction orthogonal to the ray of sight. With the help of Eqs. (\ref{rdot})-(\ref{Phidot}), it is easy to express the two coordinates $\zeta_{1}$ and $\zeta_{2}$ in the following form:
\begin{eqnarray}
 \zeta_{1}&=&r_{o}^{2}\sin\vartheta_{o}\frac{d\phi}{dr}\bigg|_{r,\;r_{o}\rightarrow\infty}
           =(L+2n\cos\vartheta_{o})\sin^{-1}\vartheta_{o},\\
 \zeta_{2}&=&r_{o}^{2}\frac{d\vartheta}{dr}\bigg|_{r,\;r_{o}\rightarrow \infty}
           =h\sin\vartheta_{o},
\end{eqnarray}
where we have taken the choice of $E=1$. Considering the quasi-equatorial case (i.e., $\vartheta_{o}=\pi/2-\psi_{o}$ with small $\psi_{o}$) and $\zeta_{1}=u$, we can obtain the orbital angular momentum $L$ and $\emph{Carter}$ constant $\mathcal{K}$, in terms of $u$ and $\psi_{o}$:
\begin{eqnarray}
 L          &=& u\cos\psi_{o}-2n\sin\psi_{o},\label{angular}\\
 \mathcal{K}&=&h^{2}\cos^{2}\psi_{o}+(u^{2}-a^{2})\sin^{2}\psi_{o}-4n\sin\psi_{o}(a-u\cos\psi_{o}+n\sin\psi_{o}). \label{cater}
\end{eqnarray}
It is clear that, when $n=0$, the result will reduce to the Kerr black hole case.

\section{Equatorial black hole lensing}
\label{SLKTN}

We devote this section to study the equatorial lensing by the rotating KTN black hole. The effects of the NUT charge $n$ and spin $a$ on it will also be investigated.

\subsection{Deflection angle}
\label{equatorDeflection}

In this subsection, we would like to consider the strong field lensing by the KTN black hole for the case that both the observer and the source lie in the equatorial plane ($\theta=\pi/2$) of the KTN black hole and the whole trajectory of the photon is also limited on this plane. Here, we can adimensionalize the metric (\ref{KTNmetric}) in terms of the Schwarzschild radii $2M$ by defining
\begin{eqnarray}
 t\rightarrow t/2M,\; r\rightarrow r/2M,\;
 a\rightarrow a/2M,\; n\rightarrow n/2M.
\end{eqnarray}
Then, the metric is reduced to
\begin{eqnarray}
 ds^{2}=-A(r)dt^{2}+B(r)dr^{2}+C(r)d\phi^{2}-D(r)dtd\phi, \label{newmetric}
\end{eqnarray}
with the metric coefficients given by
\begin{eqnarray}
 A(r)&=&\frac{r^{2}-r-n^{2}}{r^{2}+n^{2}},\\
 B(r)&=&\frac{r^{2}+n^{2}}{r(r-1)+a^{2}-n^{2}},\\
 C(r)&=&\frac{(r^{2}+n^{2})^{2}+a^{2}(r^{2}+r+3n^{2})}{r^{2}+n^{2}},\\
 D(r)&=&\frac{2a(r+2n^{2})}{r^{2}+n^{2}}.
\end{eqnarray}
In the equatorial plane, we can express the first-order geodesic equations (\ref{rdot})-(\ref{Phidot}) for the photon $(m^{2}=0)$ in terms of the metric coefficients $A(r)$, $B(r)$, $C(r)$ and $D(r)$, which read
\begin{eqnarray}
 \dot{t}&=&\frac{4CE-2DL}{4AC+D^{2}},\label{ttdot}\\
 \dot{r}&=&\pm 2\sqrt{\frac{CE^{2}-DEL-AL^{2}}{B(4AC+D^{2})}},\label{rrdot}\\
 \dot{\theta}&=&0,\\
 \dot{\phi}&=&\frac{2DE+4AL}{4AC+D^{2}}.\label{ppdot}
\end{eqnarray}
For simplicity, we take the choice $E=1$. On the other hand, we could rewrite (\ref{rrdot}) as
\begin{eqnarray}
 \dot{r}^{2}+V_{\textrm{eff}}=0,
\end{eqnarray}
where the effective potential reads
\begin{eqnarray}
 V_{\textrm{eff}}&=&-4\bigg(\frac{C-DL-AL^{2}}{B(4AC+D^{2})}\bigg)\nonumber\\
        &=&\frac{2aL(2n^{2}+r)-(n^{2}+r-r^{2})L^{2}-a^{2}(3n^{2}+r+r^{2})}
          {(n^{2}+r^{2})^{2}}-1.\label{veff}
\end{eqnarray}
At the minimum distance $r_{0}$ of photon trajectory, where $V_{\textrm{eff}}=0$, one could get \cite{Bozza67}
\begin{eqnarray}
 L= u&=&\frac{-D_{0}+\sqrt{4A_{0}C_{0}+D_{0}^{2}}}{2A_{0}} \nonumber\\
    &=&\frac{-a(r_{0}+2n^{2})+(r_{0}+n^{2})\sqrt{r_{0}(r_{0}-1)+a^{2}-n^{2}}}
                 {r_{0}^{2}-r_{0}-n^{2}},
\end{eqnarray}
where we have used Eq. (\ref{angular}). The subscript ``0" represents that the metric coefficients are evaluated at $r_{0}$.
Here, the sign before the square root has been chosen to be positive. Thus, a prograde photon orbit is related to $a>0$, and a retrograde one is related to $a<0$.

With the expressions (\ref{rrdot}) and (\ref{ppdot}), the deflection angle for the photon coming from infinity can be written as
\begin{eqnarray}
 \alpha(r_{0})=\phi(r_{0})-T_{\phi}/2. \label{angle}
\end{eqnarray}
Here, $T_{\phi}$ denotes the period of the angular coordinate $\phi$. As we know, there exists a conical angle on its axis of symmetry. However, for small $n$, we have $T_{\phi}\sim 2\pi$ for approximation. Thus the total azimuthal angle $\phi(r_{0})$ is given by
\begin{eqnarray}
 \phi(r_{0})=2\int_{r_{0}}^{\infty}
     \frac{\sqrt{B|A_{0}|}(D+2 LA)}
     {\sqrt{4AC+D^{2}}\sqrt{\textrm{sgn}(A_{0})\big(CA_{0}-AC_{0}+L(AD_{0}-A_{0}D)\big)}}dr,
\end{eqnarray}
where, $\textrm{sgn}(X)$ denotes the sign of $X$. Note that the metric coefficient $A_{0}$ changes its sign when $r_{0}<r_{\textrm{es}}$ and we have taken this into consideration. With a detailed examination, we can see that the deflection angle $\alpha(r_{0})$ increases with the decreasing of the parameter $r_{0}$. At a certain value of $r_{0}$, we may have $\alpha(r_{0})=2\pi$, which means that the light ray will make a complete loop around the black hole before reaching the observer. Let $r_{0}$ decrease further, the light ray may make more than one complete loop, and when $r_{0}$ approaches the photon circle radius $r_{\textrm{c}}$ (we will explain it later), the deflection angle $\alpha(r_{0})$ will be unboundedly large and the photons will be captured by the black hole.

Following the method developed by Bozza \cite{Bozza}, we can find the behavior of the deflection angle when photons get very close to the radius $r_{\textrm{c}}$. We first define two new variables $y$ and $z$
\begin{eqnarray}
 y&=&A(r),\\
 z&=&\frac{y-y_{0}}{1-y_{0}},
\end{eqnarray}
where $y_{0}=A_{0}$. With the two new variables, we can express the total azimuthal angle as
\begin{eqnarray}
 \phi(r_{0})=\int_{0}^{1}R(z,r_{0})f(z,r_{0})dz, \label{phiintegral}
\end{eqnarray}
with
\begin{eqnarray}
 R(z,r_{0})&=&\frac{2(1-y_{0})}{A'}\frac{\sqrt{B|A_{0}|}(D+2LA)}{\sqrt{4AC^{2}+CD^{2}}},
    \label{Rzr}\\
 f(z,r_{0})&=&\frac{1}{\sqrt{\frac{\textrm{sgn}(A_{0})}{C}
                        \big[CA_{0}-AC_{0}+L(AD_{0}-A_{0}D)\big]}}.\label{fzr}
\end{eqnarray}
These metric coefficients without the subscript ``0" are evaluated at $r=A^{-1}\big((1-y_{0})z+y_{0}\big)$. Note that the function $R(z,r_{0})$ is regular for all values of $z$ and $r_{0}$, while $f(z,r_{0})$ diverges at $z=0$. Thus, the integral (\ref{phiintegral}) can be separated into two parts
\begin{eqnarray}
 \phi(r_{0})=\phi_{R}(r_{0})+\phi_{D}(r_{0}),
\end{eqnarray}
with the divergent part
\begin{eqnarray}
 \phi_{D}(r_{0})=\int_{0}^{1}R(0,x_{\textrm{c}})f_{0}(z,r_{0})dz,
\end{eqnarray}
and the regular part
\begin{eqnarray}
 \phi_{R}(r_{0})=\int_{0}^{1}g(z,r_{0})dz,
\end{eqnarray}
with $g(z,r_{0})=R(z,r_{0})f(z,r_{0})-R(0,r_{\textrm{c}})f_{0}(z,r_{0})$.
In order to find the divergence of the integrand, we expand the argument of the square root of $f(z,r_{0})$ to second order in $z$ and the function $f_{0}(z,r_{0})$ is
\begin{eqnarray}
 f_{0}(z,r_{0})=\frac{1}{\sqrt{p z+q z^{2}+\mathcal{O}(z^{3})}},
\end{eqnarray}
where
\begin{eqnarray}
 p=&&\textrm{sgn}(A_{0})\frac{(1-A_{0})}{A'_{0}C_{0}}
         \bigg(A_{0}C'_{0}-A'_{0}C_{0}+L(A'_{0}D_{0}-A_{0}D'_{0})\bigg),\\
 q=&&\textrm{sgn}(A_{0})\frac{(1-A_{0})^{2}}{2C_{0}^{2}A'^{3}_{0}}
       \bigg(2C_{0}C'_{0}A'^{2}_{0}+(C_{0}C''_{0}-2C'^{2}_{0})A_{0}A'_{0}
             -C_{0}C'_{0}A_{0}A''_{0}\nonumber\\
          &&+L\big[A_{0}C_{0}(A''_{0}D'_{0}-A'_{0}D''_{0})
              +2A'_{0}C'_{0}(A_{0}D'_{0}-A'_{0}D_{0})\big]\bigg).
\end{eqnarray}
When the coefficient $p$ vanishes and the leading term of the divergence in $f_{0}$ is $z^{-1}$, we will obtain an unlimited deflection angle. Here let us give some notes on the photon sphere (for $a=0$) or photon circle (for $a\neq 0$). As we know that, for a static spherically symmetric spacetime, there exist several definitions of photon sphere \cite{Virbhadra,Claudel}. One definition states that it is a timelike hypersurface if the Einstein bending angle of a light ray is unlimited when the closest distance of approach coincides with the photon sphere. Extending this case to the axisymmetric spacetime, there will be a photon circle rather than photon sphere in the equatorial plane.
When the closest distance of photon orbit coincides with this photon circle radius $r_{\textrm{c}}$, we will get an unlimited deflection angle. Therefore, solving $p=0$, we will obtain the radius $r_{\textrm{c}}$ of the orbit, which is determined by the following equation (for a detailed definition of photon sphere in a stationary and axisymmetric spacetime, we refer readers to \cite{Perlick2})
\begin{eqnarray}
 A_{\textrm{c}}C'_{\textrm{c}}-A'_{\textrm{c}}C_{\textrm{c}}
    +L_{\textrm{c}}(A'_{\textrm{c}}D_{\textrm{c}}-A_{\textrm{c}}D'_{\textrm{c}})=0.
    \label{photonsphere}
\end{eqnarray}
Here, the prime indicates the derivative with respect to $r$ and the subscript ``c" represents that the metric coefficients are evaluated at $r=r_{\textrm{c}}$. In fact, we can obtain the photon circle radius $r_{\textrm{c}}$ through the effective potential $V_{\textrm{eff}}$ (\ref{veff}) by solving $V_{\textrm{eff}}=0$ and $\frac{\partial V_{\textrm{eff}}}{\partial r}=0$.

\begin{figure*}
\centerline{
\includegraphics[width=10cm,height=8cm]{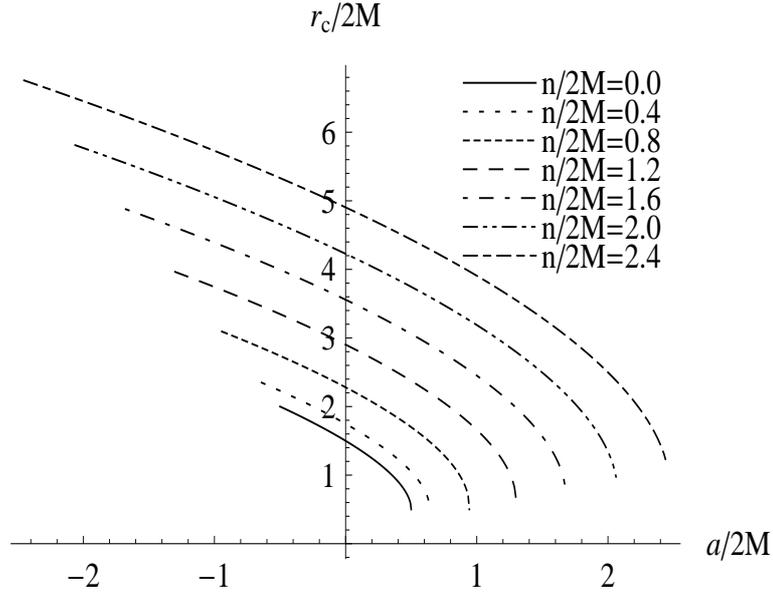}}
\caption{The radius $r_{\textrm{c}}$ as a function of the spin $a$ for the Kerr black hole (black solid line) and the KTN black hole.}\label{PRS}
\end{figure*}

\begin{figure*}
\centerline{
\includegraphics[width=10cm,height=8cm]{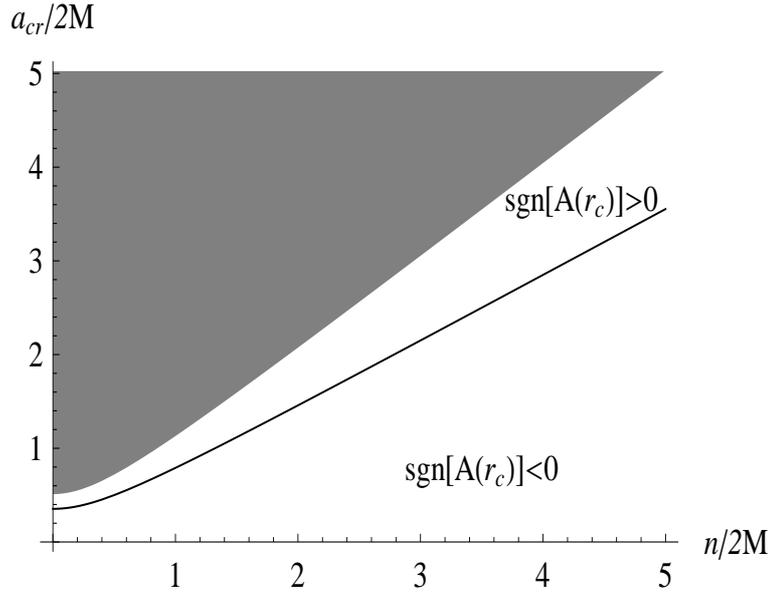}}
\caption{The critical spin $a_{\textrm{cr}}$ vs the NUT charge $n$. The shadow area denotes the case of naked singularity.}\label{PACR}
\end{figure*}

Here for the KTN black hole case, the radius $r_{\textrm{c}}$ is determined by the following equation
\begin{eqnarray}
 4r_{\textrm{c}}^{6}&-&12r_{\textrm{c}}^{5}+3(3-8n^{2})r_{\textrm{c}}^{4}
   -8(a^{2}-5n^{2})r_{\textrm{c}}^{3}\nonumber\\
   &+&2n^{2}(18n^{2}-16a^{2}-3)r_{\textrm{c}}^{2}
   +4n^{2}(2a^{2}-3n^{2})r_{\textrm{c}}+n^{4}=0.
\end{eqnarray}
Obviously, this equation is of 6th degree in the radius $r_{\textrm{c}}$. So, for the fixed values of $a$ and $n$, we will obtain six values for $r_{\textrm{c}}$. However, there are always two imaginary values and two negatives, which are unphysical. And the two remaining ones are larger than $r_{\textrm{h}}$. From the physical viewpoint, the larger one of the two is the radius of the retrograde circular photon orbit with $a<0$ and the smaller one is that of the prograde circular photon orbit with $a>0$. Thus, consider that $|a|=a$ and $|a|=-a$ together, $r_{\textrm{c}}$ is computed numerically and is plotted in Fig. \ref{PRS} for different values of $n$. The black solid line with $n=0$ describes the case of the Kerr black hole, which has a analytical expression $r_{\textrm{c}}=1+\cos(\frac{2}{3}\arccos(\pm 2|a|))$ corresponding to the radius of the prograde and retrograde circular photon orbits, respectively. For $n=0$ and $a=0$, we get the radius of photon sphere $r_{\textrm{ps}}=\frac{3}{2}$ for the Schwarzschild black hole. We can see that, for a fixed spin $a$, the radius $r_{\textrm{c}}$ increases with the NUT charge $n$. Note that the light ray trajectory is a straight line at infinity in equatorial plane, so the value of $n/2M$ could take a large value for this equatorial black hole lensing. It is also clear that, for a positive $a$, photons are allowed to get closer to the black hole, and enter even the ergosphere at some values of the spin $a$. We calculate the critical spin $a_{\textrm{cr}}$, where the circular radius $r_{\textrm{c}}$ coincides with the ergosphere in the equatorial plane
\begin{eqnarray}
 a_{\textrm{cr}}=\frac{\sqrt{1+4n^{2}}}{2\sqrt{2}}.
\end{eqnarray}
It reduces to the Kerr black hole case $a_{\textrm{cr}}=\frac{1}{2\sqrt{2}}$ for $n=0$. The behavior of the critical spin $a_{\textrm{cr}}$ is shown in Fig. \ref{PACR}. The shadow area on the top of Fig. \ref{PACR} represents the case of naked singularity. The remaining area is divided into two regions by the critical curve. Below the curve, we have sgn$(A_{\textrm{c}})=1$ and sgn$(A_{\textrm{c}})=-1$ for another region. It is also clear that, from Fig. \ref{PRS}, the prograde photons always have smaller values of $r_{\textrm{c}}$ than the retrograde ones. So, we may conclude that the retrograde photons are captured more easily than the prograde ones. It is also clear that, for a fixed spin $a$, the Kerr black hole has the minimum value of the radius $r_{\textrm{c}}$, which implies that the photons are more easily captured by the KTN black hole than by the Kerr black hole.


For the case $r_{0}\sim r_{\textrm{c}}$, the deflection angle can be expanded in the following form \cite{Bozza}
\begin{eqnarray}
 \alpha(u)=-\bar{a}\log\big(\frac{u}{u_{\textrm{c}}}-1\big)
                +\bar{b}+\mathcal{O}(u-u_{\textrm{c}}).\label{Atheta}
\end{eqnarray}
The coefficients $u_{\textrm{c}}$, $\bar{a}$ and $\bar{b}$ are given by
\begin{eqnarray}
 u_{\textrm{c}}&=&L|_{r_{0}=r_{\textrm{c}}},\\
 \bar{a}&=&\frac{R(0,r_{\textrm{c}})}{2\sqrt{q_{\textrm{c}}}}
            =\sqrt{\frac{2A_{\textrm{c}}B_{\textrm{c}}}
                   {A_{\textrm{c}}C_{\textrm{c}}''-A_{\textrm{c}}''C_{\textrm{c}}+
                   u_{\textrm{c}}(A_{\textrm{c}}''D_{\textrm{c}}
                       -A_{\textrm{c}}D_{\textrm{c}}'')}},\\
 \bar{b}&=&-\pi+b_{R}+\bar{a}\log\bigg(\frac{4q_{\textrm{c}}C_{\textrm{c}}}
             {u_{\textrm{c}}|A_{\textrm{c}}|(D_{\textrm{c}}
                      +2u_{\textrm{c}}A_{\textrm{c}})}\bigg),
\end{eqnarray}
where $q_{\textrm{c}}=q|_{r=r_{\textrm{c}}}$. In order to obtain the coefficient $b_{R}$, we expand $\phi_{R}(r_{0})$ at $r_{\textrm{c}}$
\begin{eqnarray}
 \phi_{R}(r_{0})=\sum_{n=0}^{\infty}\frac{1}{n!}(r_{0}-r_{\textrm{c}})^{n}
                      \int_{0}^{1}\frac{\partial^{n}g}
                        {\partial r_{0}^{n}}\bigg|_{r_{0}=r_{\textrm{c}}}dz.
\end{eqnarray}
Thus, we get
\begin{eqnarray}
 b_{R}=\phi_{R}(r_{\textrm{c}})
      =\int_{0}^{1}g(z,r_{\textrm{c}})dz,
\end{eqnarray}
which can be obtained numerically.

\begin{figure*}
\centerline{
\includegraphics[width=10cm,height=8cm]{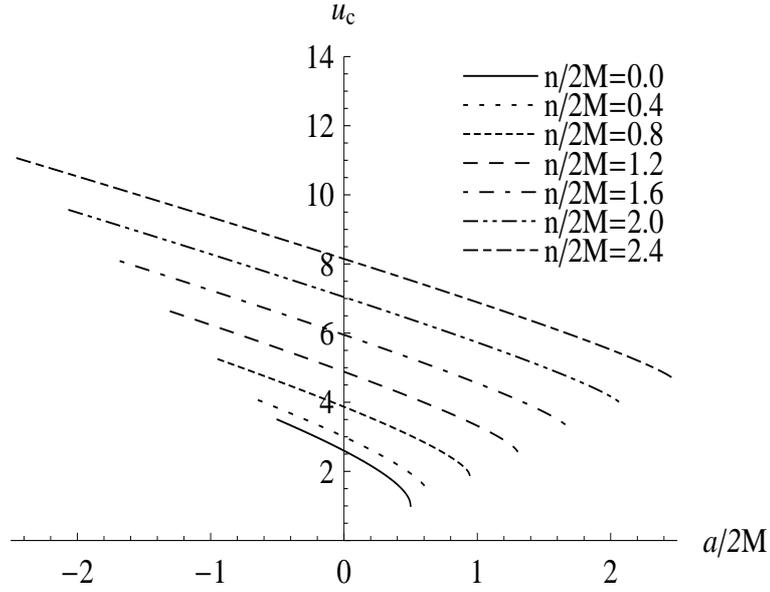}}
\caption{Variation of the minimum impact parameter $u_{\textrm{c}}$ with the spin $a$ of the Kerr black hole (black solid line) and the KTN black hole.}\label{PUPS}
\end{figure*}

\begin{figure*}
\centerline{
\includegraphics[width=10cm,height=8cm]{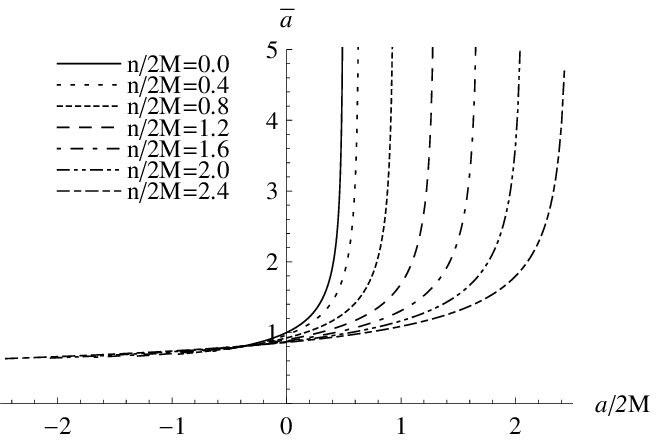}}
\caption{Strong deflection limit coefficient $\bar{a}$ as a function of the spin $a$ of the Kerr black hole (black solid line) and the KTN black hole.}\label{PABAR}
\end{figure*}
\begin{figure*}
\centerline{
\includegraphics[width=10cm,height=8cm]{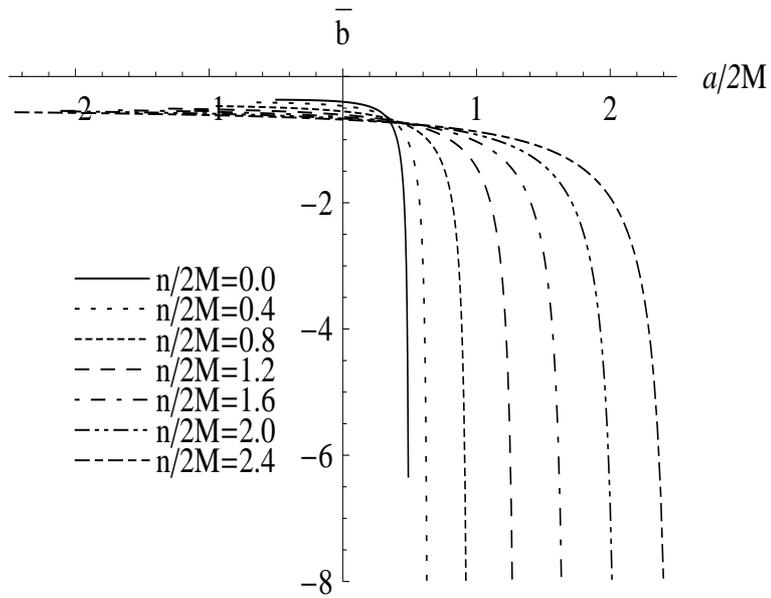}}
\caption{Strong deflection limit coefficient $\bar{b}$ as a function of the spin $a$ of the Kerr black hole (black solid line) and the KTN black hole.}\label{PBBAR}
\end{figure*}

Fig. \ref{PUPS} depicts the minimum impact parameter $u_{\textrm{c}}$ as a function of the spin $a$ for different values of the NUT charge $n$. We can see that $u_{\textrm{c}}$ has the similar behavior as $r_{\textrm{c}}$. We also can see that the Kerr black hole has smaller value of $u_{\textrm{c}}$ for a fixed spin $a$ than that of the KTN black hole. The coefficients of the strong deflection limit $\bar{a}$ and $\bar{b}$ are illustrated in Figs. \ref{PABAR} and \ref{PBBAR}. It is easy to obtain that, for a fixed NUT charge $n$, $\bar{a}$ grows with the spin $a$, while $\bar{b}$ decreases. And both the coefficients diverge at $a=\frac{1}{2}\sqrt{1+4n^{2}}$, which corresponds to the extremal black hole. The divergence of the coefficients implies that, in the strong field limit, the deflection angle no longer represents a reliable result.

\begin{figure*}
\centerline{
\includegraphics[width=10cm,height=8cm]{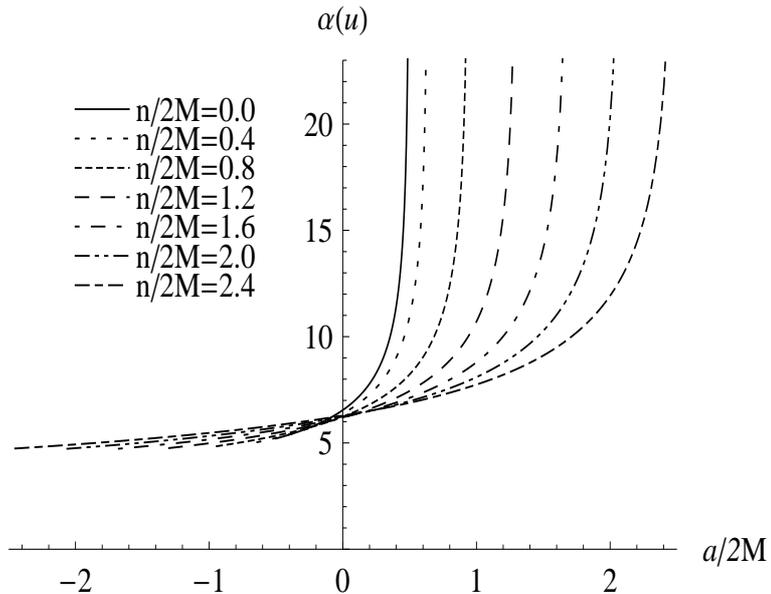}}
\caption{The deflection angle $\alpha(u)$ vs the spin $a$ of the Kerr black hole (black solid line) and the KTN black hole for $u=u_{\textrm{c}}+0.0025$.}\label{PA}
\end{figure*}

With $\bar{a}$ and $\bar{b}$, we can obtain the deflection angle $\alpha(u)$. The change of $\alpha(u)$ with spin $a$ for $u=u_{\textrm{c}}+0.0025$ is displayed in Fig. \ref{PA}. From it, we obtain the results: (1) For the Kerr black hole (described by the black solid line in Fig. \ref{PA}), $\alpha(u)$ monotonically increases with the spin $a$, and $\alpha(u)$ of prograde photon is larger than the retrograde one. (2) For the KTN black hole with the NUT charge $n\neq 0$, the change of the deflection angle $\alpha(u)$ with the spin $a$ has the same behavior as the Kerr black hole. (3) For a fixed value of the spin $a$ less than some negative value, the deflection angle $\alpha(u)$ of the Kerr black hole is smaller than that of the KTN black hole; and for a spin $a$ greater than this negative value, $\alpha(u)$ of the Kerr black hole is larger than that of the KTN black hole.

\subsection{Lens equation}
\label{lensequation}

In order to describe the lensing, several lens equations were introduced (such as \cite{Ohanian,Virbhadra337,Dabrowski535}). These equations principally differ from each other for the different choices of the variables. In \cite{Bozza103005}, the author gave a detailed comparison of each lens equation and he suggested that the Ohanian lens equation is the best approximate lens equation. So, in this subsection, we would like to give a brief introduction to the equatorial lens equation. General, the optical axis is defined as the line joining the observer and the lens. Setting the black hole in the origin, we denote the angle between the direction of the source and the optical axis by $\gamma$. Then the case $\gamma=0$ corresponds to that the source, lens and observer are perfectly aligned. From the lensing geometry, the angle $\gamma$ can be expressed as
\begin{eqnarray}
 \gamma=-\alpha(\theta)+\theta+\overline{\theta}\;\textrm{mod}\; 2\pi,\label{equatorialequation}
\end{eqnarray}
where the impact angle is
\begin{eqnarray}
 \overline{\theta}\simeq \frac{u}{D_{\textrm{LS}}}
                  \simeq \frac{\theta D_{\textrm{OL}}}{D_{\textrm{LS}}}.
\end{eqnarray}
Although the spacetime has a conical singularity, the lens equation (\ref{equatorialequation}) is still held, because that all azimuth angles are measured in coordinate $\phi$ with period $T_{\phi}=2\pi$.
$D_{\textrm{LS}}$ measures the distance between the lens and source, $D_{\textrm{OL}}$ for the observer and lens, and $D_{\textrm{OS}}$ for the observer and the source \cite{Bozza103005}.
A usual relation between them $D_{\textrm{OS}}=D_{\textrm{OL}}+D_{\textrm{LS}}$ is held.
The lens equation can be reexpressed as
\begin{eqnarray}
 \gamma=\frac{D_{\textrm{OL}}+D_{\textrm{LS}}}{D_{\textrm{LS}}}\theta-\alpha(\theta)
                  \;\textrm{mod}\;2\pi.\label{lensgamma}
\end{eqnarray}
Since the angle $\gamma\in [-\pi,\;\pi]$, the source and the observer could be on the opposite side or on the same side of the lens.

Consider that $\theta=u/D_{\textrm{OL}}\ll 1$, we have $\gamma\simeq-\alpha(\theta)$. Moreover, one may find
\begin{eqnarray}
 \theta_{n}^{0}=\frac{u_{\textrm{c}}}{D_{\textrm{OL}}}(1+e_{n}),
\end{eqnarray}
where $e_{n}=e^{(\bar{b}+\gamma-2n\pi)/\bar{a}}$ and $n$ is the number of loops done by the photon around the black hole. Expanding the deflection angle $\alpha(\theta)$ around $\theta_{n}^{0}$ to the first order, we get
\begin{eqnarray}
 \alpha(\theta)&=&\alpha(\theta_{n}^{0})
                +\frac{\partial\alpha}{\partial\theta}\big|_{\theta_{n}^{0}}
                  (\theta-\theta_{n}^{0})
                  +\mathcal{O}(\theta-\theta_{n}^{0})^{2}\;\textrm{mod}\;2\pi\nonumber\\
               &\simeq& -\gamma-\frac{\bar{a}D_{\textrm{OL}}}{u{\textrm{c}}}
                  (\theta-\theta_{n}^{0})
                    +\mathcal{O}(\theta-\theta_{n}^{0})^{2}\;\textrm{mod}\;2\pi.
\end{eqnarray}
Neglecting the higher order terms and plugging this result into the equatorial lens equation (\ref{lensgamma}), we obtain the position of the $n$-th relativistic image
\begin{eqnarray}
 \theta_{n}\simeq\theta_{n}^{0}\bigg(1-
                \frac{u_{\textrm{c}}e_{n}(D_{\textrm{OL}}+D_{\textrm{LS}})}
                  {\bar{a}D_{\textrm{OL}}D_{\textrm{LS}}}\bigg).\label{thetan}
\end{eqnarray}
From (\ref{thetan}), it is easy to find that the correction is much smaller than $\theta_{n}^{0}$ for $D_{\textrm{OL}}\sim D_{\textrm{LS}}\gg 1$. 

\section{Quasi-equatorial black hole lensing}
\label{QLKTN}

It is known that, in an axisymmetric spacetime, even if the central ray of the light bundle is in the equatorial plane, most of the rays in the bundle will leave the equatorial plane. Thus, in this section, we would like to investigate the quasi-equatorial lensing by the KTN black hole. It was suggested by Bozza \cite{Bozza67} that, in order to study the quasi-equatorial lensing, a two-dimensional lens equation is needed. Accordingly, an additional parameter $\psi=\pi/2-\vartheta$, which measures the declination between the orbit plane of the light ray and the equatorial plane, should be included in. It is also worth to mention that, we here adopt small $n/2M$ approximation and $\psi\sim 0$, so the technique for the Kerr black hole case is also applicable for this case.

\subsection{Precession of the orbit}

\begin{figure*}
\centerline{
\includegraphics[width=10cm,height=8cm]{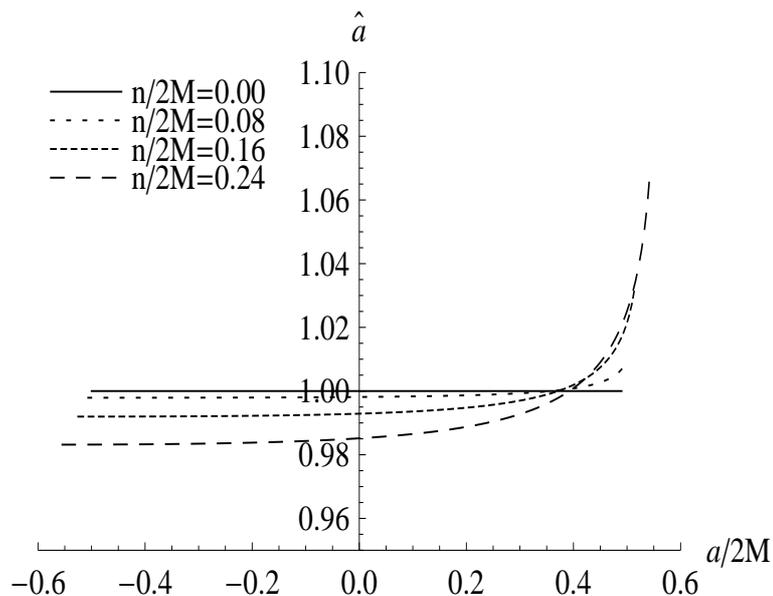}}
\caption{The coefficient $\hat{a}$ vs the spin $a$ for different values of the NUT charge $n$.}\label{PAHAT}
\end{figure*}

\begin{figure*}
\centerline{
\includegraphics[width=10cm,height=8cm]{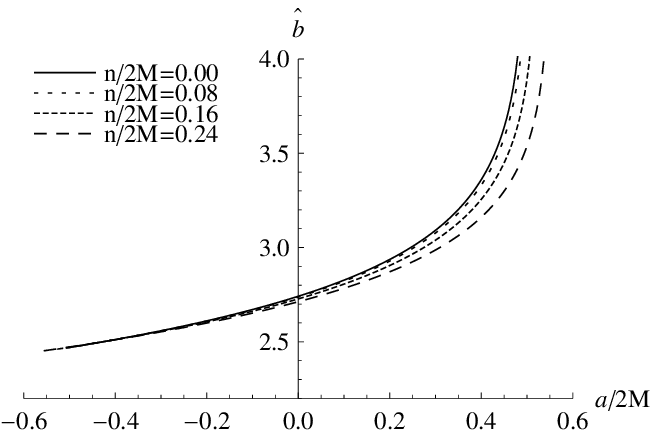}}
\caption{The coefficient $\hat{b}$ vs the spin $a$ for different values of the NUT charge $n$.}\label{PBHAT}
\end{figure*}

\begin{figure*}
\centerline{
\includegraphics[width=10cm,height=8cm]{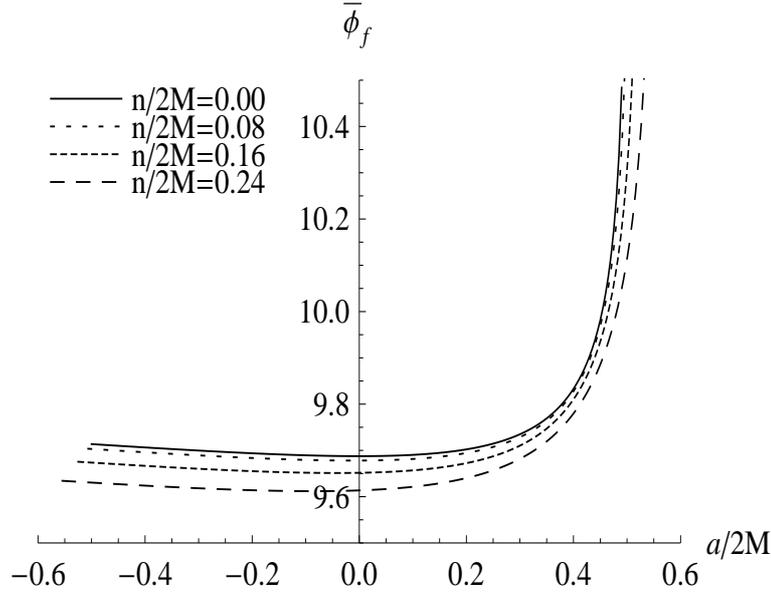}}
\caption{Variation of the values of $\bar{\phi}_{f}$ with the spin $a$ for fixed $u=u_{\textrm{c}}+0.0025$.}\label{PPHI}
\end{figure*}

The precession of the orbit at small declination is an important ingredient for the study of quasi-equatorial lensing. Here, we only consider the small $n$ case. Similar to the Kerr black hole, the evolution equation for $\psi$ as a function of the azimuthal angle $\phi$ for the KTN black hole approximatively reads
\begin{eqnarray}
 \frac{d\psi}{d\phi}\simeq\pm \omega(\phi)\sqrt{\bar{\psi}^{2}-\psi^{2}},\label{psiphi}
\end{eqnarray}
where
\begin{eqnarray}
 \bar{u}&=&\sqrt{u^{2}-a^{2}},\\
 \bar{\psi}&=&\sqrt{\frac{h^{2}}{\bar{u}^{2}}+\psi_{0}^{2}},\label{phibar}\\
 \omega(\phi)&=&\bar{u}\frac{a^{2}-n^{2}+r(r-1)}{a(r+2n^{2})-L(n^{2}+r-r^{2})}.
             \label{omega}
\end{eqnarray}
It is obvious that the parameter $\omega(\phi)$ depends on the NUT charge $n$ and spin $a$. When $n=0$, $\omega(\phi)$ will reduce to the Kerr black hole case \cite{Bozza67}. The solution of (\ref{psiphi}) is
\begin{eqnarray}
 \psi(\phi)=\bar{\psi}\cos(\bar{\phi}+\phi_{0}), \label{phi0}
\end{eqnarray}
with $\bar{\phi}=\int^{\phi}_{0}\omega(\phi')d\phi'$ and $\phi_{0}$ a constant. Since the photon comes from infinity and returns to infinity, the deflection angle is
\begin{eqnarray}
 \bar{\phi}_{f}=2\int_{r_{0}}^{\infty}\omega(r)\frac{d\phi}{dr}dr
               =\int_{0}^{1}R_{\omega}(z,r_{0})f(z,r_{0})dz,
\end{eqnarray}
where
\begin{eqnarray}
 R_{\omega}(z,r_{0})=\omega(r)R(z,r_{0}).
\end{eqnarray}
The quantities $R(z, r_{0})$ and $f(z, r_{0})$ are given by Eqs. (\ref{Rzr}) and (\ref{fzr}), respectively. Since there are no singularities in $\omega(r)$, it can be absorbed into the regular function $R(z, r_{0})$. As a result, we can apply the same technique used in section \ref{SLKTN} for this case. Thus, the representation of $\bar{\phi}_{f}$ in the strong field limit reads
\begin{eqnarray}
 \bar{\phi}_{f}&=&-\hat{a}\ln\bigg(\frac{u}{u_{\textrm{c}}}-1\bigg)+\hat{b},\label{phiangle}\\
 \hat{a}&=&\frac{R_{\omega}(0,r_{\textrm{c}})}{2\sqrt{q_{\textrm{c}}}},\\
 \hat{b}&=&\hat{b}_{R}
       +\hat{a}\ln\bigg(\frac{4q_{\textrm{c}}C_{\textrm{c}}}
           {u_{\textrm{c}}|A_{\textrm{c}}|(D_{\textrm{c}}+2u_{\textrm{c}} A_{\textrm{c}})}\bigg),
\end{eqnarray}
with
\begin{eqnarray}
 \hat{b}_{R}=\int_{0}^{1}
      \bigg[R_{\omega}(z,r_{0})f(z,r_{0})
    -R_{\omega}(0,r_{\textrm{c}})f_{0}(z,r_{0})\bigg]dz.
\end{eqnarray}
The quantities $\hat{a}$ and $\hat{b}$ are evaluated numerically at the circular radius $r_{\textrm{c}}$ and their behavior is shown in Figs. \ref{PAHAT} and \ref{PBHAT}. From them, we can find that, for the Kerr black hole, we always have $\hat{a}=1$ for all spin $a$. However, for nonvanishing charge $n$, both quantities $\hat{a}$ and $\hat{b}$ are found to increase with the spin $a$. And both of them diverge at $a=1/2$. Variation of $\bar{\phi}_{f}$ with the spin $a$ is presented in Fig. \ref{PPHI} with $u=u_{\textrm{c}}+0.0025$. For fixed charge $n$, it increases with the spin $a$. And for fixed spin $a$, it decreases with the charge $n$.

\subsection{Lensing at small declination}
\label{smalldeclinations}

In this subsection, let us study the positions and magnification of the images at small declination. Here we set the source and observer at heights $h_{S}$ and $h_{O}$, respectively. In order to ensure the small declination condition during the whole trajectory of the photon, we assume that the following relation is satisfied
\begin{eqnarray}
 (h_{O},h_{S})\ll u\ll (D_{OL},D_{LS}).
\end{eqnarray}
With the small declination hypothesis, the polar lens equation connecting the positions of the source and the observer can be expressed as \cite{Bozza67}:
\begin{eqnarray}
 h_{S}=h_{O}\bigg(\frac{D_{\textrm{OL}}}{\bar{u}}S
                       -C\bigg)-\psi_{0}
                 \bigg((D_{\textrm{OL}}+D_{\textrm{LS}})C
                  -\frac{D_{\textrm{OL}}D_{\textrm{LS}}}{\bar{u}}S\bigg),
\label{polar}
\end{eqnarray}
where $S=\sin\bar{\phi}_{f}$ and $C=\cos\bar{\phi}_{f}$. The inclination $\psi_{0}$ is related to the heights of the observer and the source. For the $n$-th image, the inclination $\psi_{0}$ and the height $h_{n}$ are
\begin{eqnarray}
 \psi_{0,n}&=&\frac{\bar{u}(h_{S}+h_{O}C_{n})-h_{O}D_{\textrm{LS}}S_{n}}
              {D_{\textrm{OL}}D_{\textrm{LS}}S_{n}
                  -\bar{u}(D_{\textrm{OL}}+D_{\textrm{LS}})C_{n}},\\
 h_{n}&=&\frac{\bar{u}(h_{S}D_{\textrm{{OL}}}-h_{O}D_{\textrm{{LS}}}C_{n})}
            {D_{\textrm{OL}}D_{\textrm{LS}}S_{n}
                  -\bar{u}(D_{\textrm{OL}}+D_{\textrm{LS}})C_{n}},
\end{eqnarray}
where $S_{n}$ and $C_{n}$ are the values of $S$ and $C$ evaluated at $\bar{\phi}_{f}=\bar{\phi}_{f,n}$. It is clear that in the neighborhood of $\bar{\phi}_{f}=k\pi$, the denominators of the $\psi_{0,n}$ and $h_{n}$ will vanish. Thus, the position of the caustic points is naturally determined by the denominators of the $\psi_{0,n}$ and $h_{n}$, i.e.,
\begin{eqnarray}
     K(\gamma)=D_{\textrm{OL}}D_{\textrm{LS}}S_{n}
                  -\bar{u}(D_{\textrm{OL}}+D_{\textrm{LS}})C_{n}. \label{KK}
\end{eqnarray}
So, the caustic points are located at $\bar{\phi}_{f}=k\pi$. Solving it, we could obtain the angular positions of the caustic points:
\begin{eqnarray}
     \gamma_{k}=-\bar{b}+\frac{\bar{a}}{\hat{a}}(\hat{b}-k\pi).
\end{eqnarray}
For each $k$, there exist one caustic point for the prograde photons and one for the retrograde photons. The case $k=1$ corresponds to the weak field caustic points and others to that in the strong field limit approximation.

Next, we would like to study the magnification of the images near the caustic points. The magnification is defined as the ratio of the angular area element of the image and that of the source without lens,
\begin{eqnarray}
  \mu=\frac{d^{2}\mathcal{A_{I}}}{d^{2}\mathcal{A}_{S}}
        =\frac{(D_{\textrm{OL}}+D_{\textrm{LS}})^{2}}{D_{\textrm{LS}}}\frac{1}{|J|}.
\end{eqnarray}
The Jacobian determinant reads
\begin{eqnarray}
 |J|=\bigg|\frac{\partial \gamma}{\partial\theta}
             \frac{\partial h_{S}}{\partial\psi_{0}}\bigg|
    =\frac{\bar{u}u_{\textrm{c}}e_{\gamma}}
    {\bar{a}D_{\textrm{OL}}|K(\gamma)|},\label{jacobian}
\end{eqnarray}
where
\begin{eqnarray}
  e_{\gamma}=e^{(\bar{b}+\gamma)/\bar{a}}.
\end{eqnarray}
The number of loops $n=\frac{\pi-\gamma}{2\pi}$ made by the photon has been coded in $\gamma$, so $\gamma$ here can take an arbitrary value. And he angular position of the source is described by $\gamma$ mod $2\pi$. It is also worth pointing out that different values of $\gamma$ differing by a multiple of $2\pi$ represent the same source position with respect to the lens but reached by photons performing a different number of loops.

\begin{figure*}
\centerline{
\includegraphics[width=10cm,height=8cm]{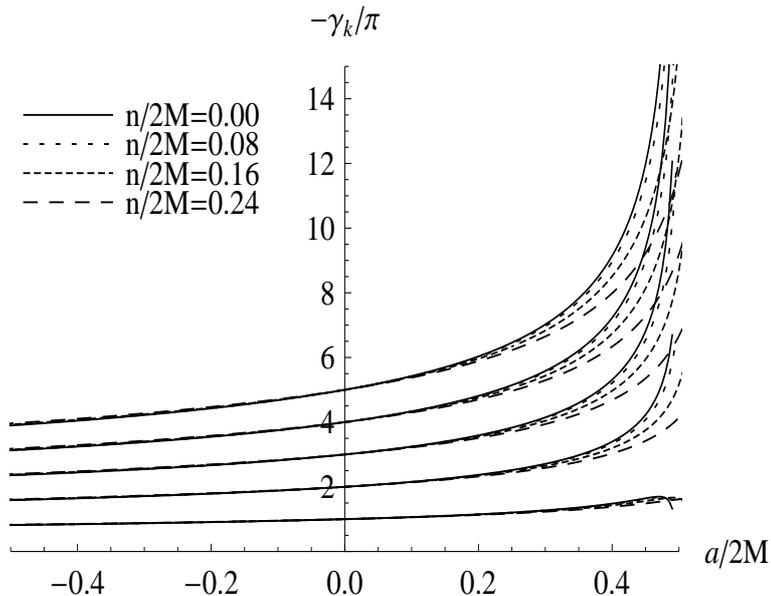}}
\caption{The angular position of the first five relativistic caustic points. From below to above $k=2,3,4,5,6$.}\label{PGAMMA}
\end{figure*}

\begin{figure}
 \includegraphics[width=8cm]{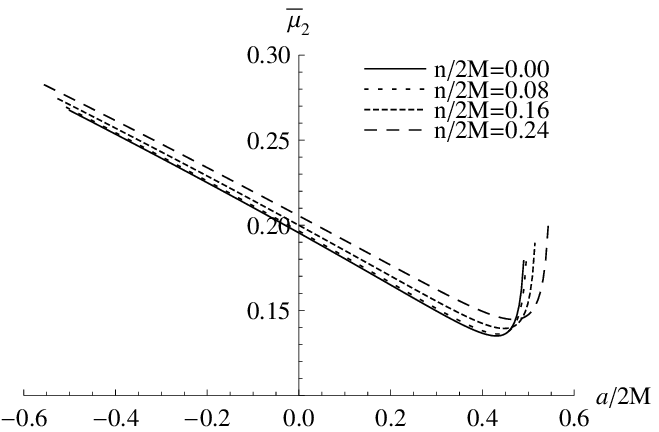}
 \includegraphics[width=8cm]{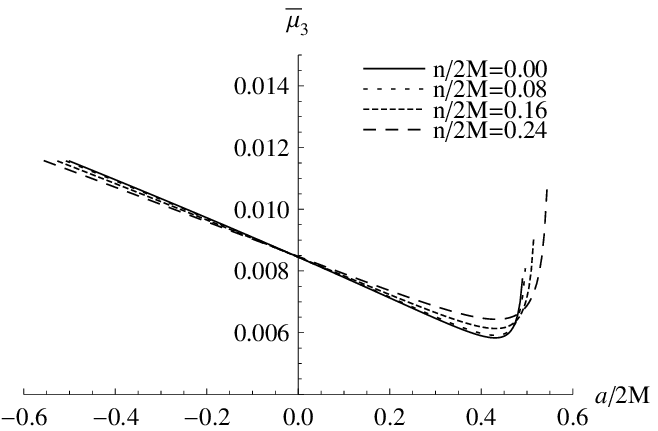}
 \includegraphics[width=8cm]{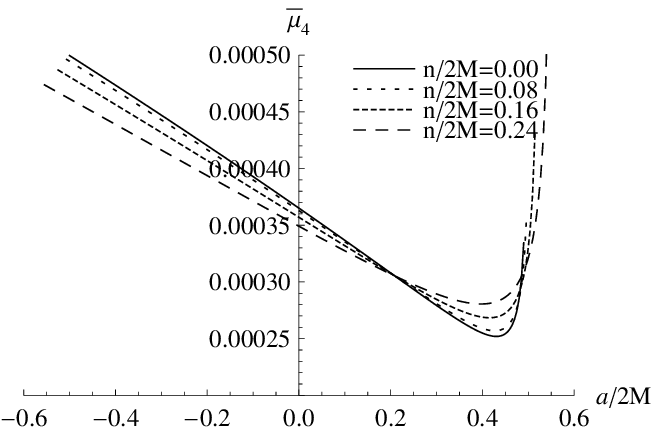}
 \includegraphics[width=8cm]{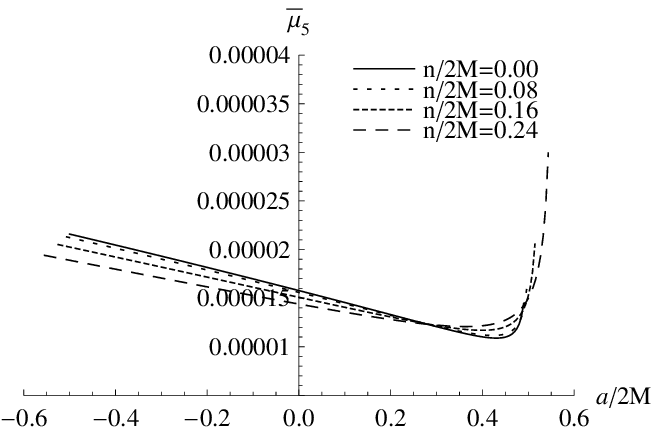}
\caption{Behaviors of the magnifying power $\bar{\mu}_{k}$ for $k=2,3,4,5$.} \label{PMU}
\end{figure}

We show the positions of the first five relativistic caustic points in Fig. \ref{PGAMMA}. When the charge $n=0$, it describes the result for the Kerr black hole and it is consistent with the result in \cite{Bozza67}. When $n=0$ and $a=0$, it recovers the result of the Schwarzschild black hole and we have $\gamma_{k}=-(k-1)\pi$, which means all the caustic points for the Schwarzschild black hole are located at the optical axis. While for the Kerr black hole with or without the NUT charge $n$, all caustic points of the source are not on the optical axis. We can also see that the caustic points are anticipated for negative spin $a$ and delayed for positive spin $a$. It is obvious that the caustic curves can move very far from the optical axis at large values of the spin $a$. Moreover, we can obtain the result that the NUT charge $n$ has strong impact on the prograde photons than the retrograde ones. From Eq. (\ref{jacobian}), we find that the Jacobian determinant will diverge in the caustic points. Thus, in order to describe the magnification of the enhanced images created by a source near the caustic points, we expand (\ref{KK}) around $\gamma_{k}$ and retain the first term
\begin{eqnarray}
 K(\gamma)\simeq
         -\frac{\hat{a}D_{\textrm{OL}}D_{\textrm{LS}}}{\bar{a}}(\gamma-\gamma_{k}(a))
         +\mathcal{O}(\gamma-\gamma_{k}(a))^{2}.
\end{eqnarray}
Then the magnification of the enhanced images is expressed as
\begin{eqnarray}
 u_{k}^{\textrm{enh}}&=&\frac{(D_{\textrm{OL}}+D_{\textrm{LS}})^{2}}
                {D_{\textrm{OL}}^{2}D_{\textrm{LS}^{2}}}
                \frac{\bar{\mu}_{k}(a)}{|\gamma-\gamma_{k}|},\\
 \bar{\mu}_{k}(a)&=&\bar{u}u_{\textrm{c}}e_{\gamma_{k}}(a)\hat{a}^{-1},\label{mu}\\
 e_{\gamma_{k}}(a)&=&e^{(\hat{b}-k\pi)/\hat{a}}.
\end{eqnarray}
The quantity $\bar{\mu}_{k}$ denotes the magnifying power for the KTN black hole close to the caustic points. The behaviors of $\bar{u}_{k}$ for $k=2, 3, 4, 5$ are shown in Fig. \ref{PMU}. From it, we can get the results that $\bar{u}_{k}$ first decreases linearly with and then grows with the spin $a$. And each $\bar{u}_{k}$ shows a divergence for the extremal black hole case. However, we should keep in mind that the result that the behavior of the magnifying power around an extremal black hole is inaccurate because the strong field limit approximation there breaks down. It is also clear that the magnification falls rapidly with the increasing of $k$.

\begin{figure*}
\centerline{
 \includegraphics[width=8cm]{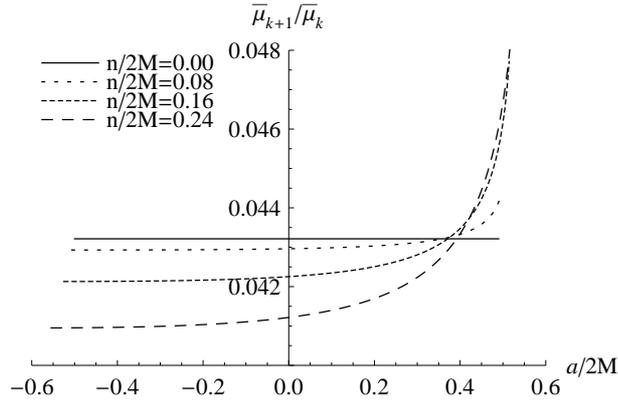}}
\caption{The ratio $\bar{\mu}_{k+1}/\bar{\mu}_{k}$ vs. the spin $a$. The black solid line has a constant value $e^{-\pi}$.} \label{PMUMU}
\end{figure*}

It is obvious that the shapes of $\bar{\mu}_{k}$ shown in Fig. \ref{PMU}, remain more or less the same for different values of $k$. And for two consecutive numbers $k$ and $k+1$, we get
\begin{eqnarray}
 \frac{\bar{\mu}_{k+1}}{\bar{\mu}_{k}}= e^{-\pi/\hat{a}}.
\end{eqnarray}
Since $\hat{a}=1$, the Kerr black hole has a constant ratio $\bar{\mu}_{k+1}/\bar{\mu}_{k}=e^{-\pi}$. The numerical results of the ratio for the NUT charge $n\neq 0$ can be found in Fig. \ref{PMUMU}. We can find that, for fixed charge $n \neq 0$, the ratio increases with the spin $a$. The ratio decreases with the charge $n$ for fixed spin $a$ with small value, and the result for fixed high spin $a$ is the reverse.

\section{Critical curves and caustic structure}
\label{Critical}

It is known that, for the Schwarzschild black hole lens, if the source, lens and the observer are strictly aligned, then a large Einstein ring and two infinite series of concentric relativistic Einstein rings will appear \cite{Bozza}. However, for the rotating black hole lens \cite{Bozza67,Gyulchev75}, the result shows a big difference. The nontrivial caustic structures appear, and the caustics drift away from the optical axis and acquire a finite extension. Especially, for a black hole with high spin $a$, only one image rather than two infinite series of relativistic images will be observed. In this section, we would like to study the effect of the NUT charge $n$ on the critical curves and the caustic structure.

\begin{figure}
 \includegraphics[width=8cm]{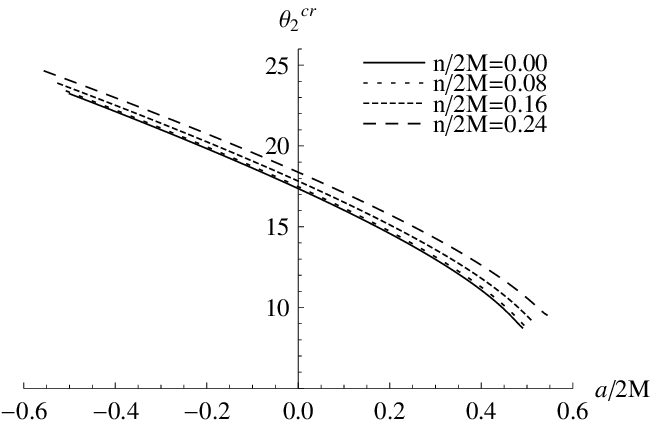}
 \includegraphics[width=8cm]{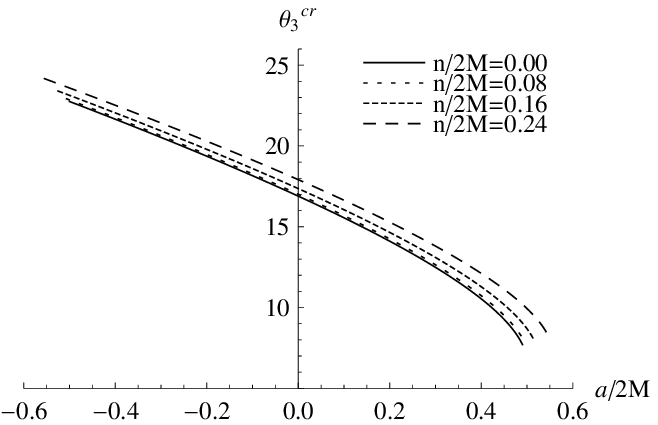}
 \includegraphics[width=8cm]{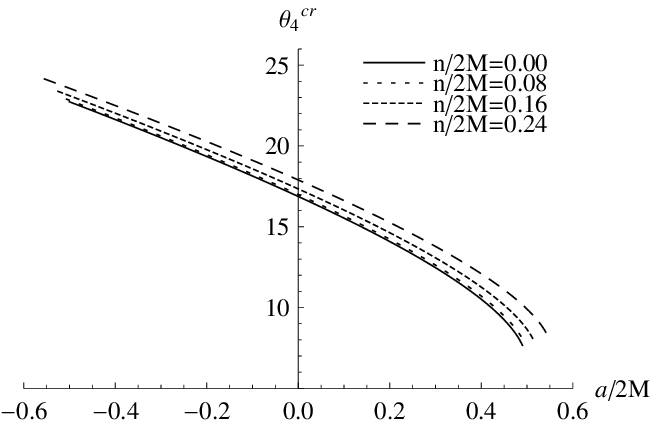}
 \includegraphics[width=8cm]{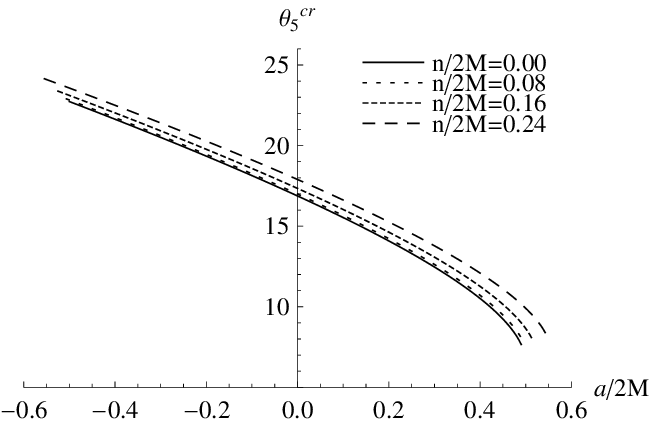}
\caption{Intersections of the strong deflection limit critical curves with the equatorial plane for $k=2,3,4,5$.} \label{PTHETA}
\end{figure}

For the KTN metric, the intersections of the critical curves with the equatorial plane in the quasi-equatorial approximation and in the strong deflection limit are
\begin{eqnarray}
 \theta_{k}^{\textrm{cr}}&\simeq&\theta_{k}^{\textrm{0,cr}}
                   \bigg(1-\frac{u_{\textrm{c}}e_{\gamma_{k}}
                  (D_{\textrm{OL}}+D_{\textrm{LS}})}
                   {\bar{a}D_{\textrm{OL}}D_{\textrm{LS}}}\bigg),
 \end{eqnarray}
 where
\begin{eqnarray}
  \theta_{k}^{\textrm{0,cr}}&=&\frac{u_{\textrm{c}}}{D_{\textrm{OL}}}(1+e_{\gamma_{k}}).
\end{eqnarray}
Here, we suppose that the gravitational field of the supermassive black hole at the center of our Milky Way can be described by the KTN metric. The mass of the supermassive black hole is estimated to be $M= 2.8\times 10^{6}M_{\odot}$, and the distances are assumed to $D_{\textrm{OL}}=8.5$ kpc, $D_{\textrm{LS}} = 1.0$ kpc. Then the numerical results for $\theta_{k}^{\textrm{cr}}$ are plotted in Fig. \ref{PTHETA}. From it, we can find that these curves are very close to each other and the critical points are close to the optical axis $\theta=0$ for positive spin $a$ and farther for  negative spin $a$.

\begin{figure*}
\center{
 \includegraphics[width=8cm]{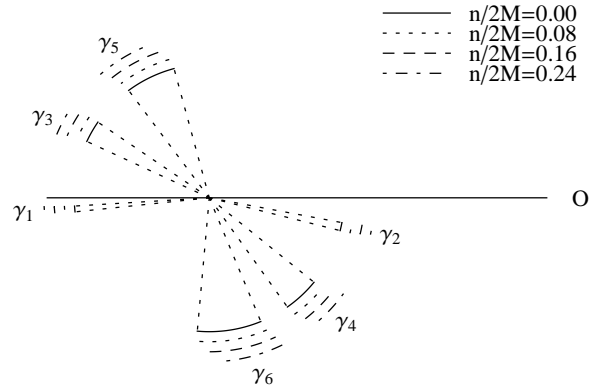}}
\caption{The first six caustics for the KTN black hole with fixed spin $a=0.12$, marked by the arcs between $\gamma_{k}(-|a|)$ and $\gamma_{k}(|a|)$.} \label{PLOCATEA}
\end{figure*}

\begin{figure*}
\center{
 \includegraphics[width=8cm]{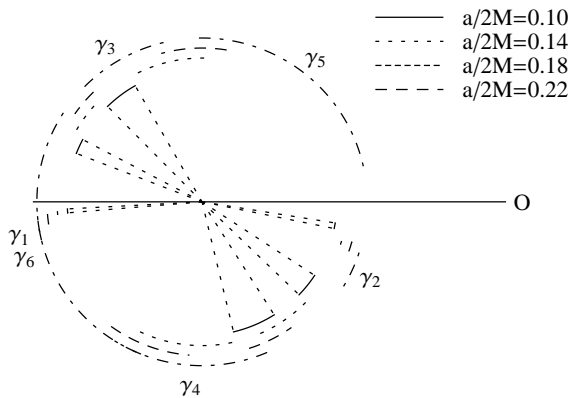}}
\caption{The first six caustics for the KTN black hole for different values of the spin $a$ with fixed charge $n=0.08$.} \label{PLOCATEB}
\end{figure*}

From above discussion, we get a result that the caustics for a rotating black hole exist a nonvanishing extension. Here we would like to study it and draw sketch maps for it. It is known that the intersection of the $k$-th caustic with the equatorial plane is determined by $\gamma_{k}(-|a|)$ and $\gamma_{k}(|a|)$. We plot the first six caustics in Fig. \ref{PLOCATEA} and Fig. \ref{PLOCATEB} for fixed spin $a$ and fixed NUT charge $n$, respectively, seen from the direction of the spin. In these figures, the centre of the circular sectors stand for the location of the black hole lens. The line connecting the centre and O (observer) is the optical axis. The circular arc denotes the angular positions $\gamma_{k}$ of the source. And the distances between the source, lens and observer are ignored in these figures. From them, we can find that the nonrelativistic caustic $\gamma_{1}$ stays close to the optical axis. And other relativistic ones drift in the anticlockwise direction for fixed spin $a$, and in the clockwise direction for fixed NUT charge $n$. In Fig. \ref{PLOCATEA}, the relativistic caustics $\gamma_{2}$, $\gamma_{4}$ and $\gamma_{6}$ are shifted to below the optical axis, and $\gamma_{3}$, $\gamma_{5}$ are shifted to above it. While, in Fig. \ref{PLOCATEB}, these relativistic caustics are shifted in the opposite direction. It is also clear that, for fixed NUT charge $n$, the caustics get larger and farther from their initial position on the optical axis when spin $a$ and $k$ increases. However, for fixed spin $a$, the caustics keep its extension and get closer from their initial position.


\section{Discussions and summary}
\label{Summary}

In this paper, we studied numerically the equatorial and quasi-equatorial lensing by the stationary, axially-symmetric KTN black hole in the strong field limit. First, we derived the first order differential system for the geodesics in the black hole background. The result shows that the NUT charge $n$ indeed influences the geodesics of particles. Then, combining with the lens equation and the null geodesics, we studied the equatorial lensing by the KTN black hole. The strong deflection limit coefficients $\bar{a}$, $\bar{b}$ and the deflection angle $\alpha(u)$ were obtained. From the numerical calculation, we found that, for a fixed NUT charge $n$, $\bar{a}$ grows with the increasing of the spin $a$, while $\bar{b}$ decreases. With the values of the strong deflection limit coefficients $\bar{a}$ and $\bar{b}$, we got the numerical value of the deflection angle. In Fig. \ref{PA}, it was plotted as a function of the spin $a$ for different values of the NUT charge $n$. From it, the Kerr black hole and the KTN black hole were found to share the same property that the deflection angle $\alpha(u)$ monotonically increases with the spin $a$ and the deflection angle of prograde photon is larger than the retrograde one. However, it was also found that, for a fixed value of spin $a$ less than a negative value, the deflection angle $\alpha(u)$ for the Kerr black hole is smaller than that for the KTN black hole. And for a spin $a$ greater than the value, $\alpha(u)$ of the Kerr black hole is larger than that of the KTN black hole.

Considering that, even if the central ray of the light bundle is in the equatorial plane, most of the rays in the bundle will leave the equatorial plane, we investigated the quasi-equatorial lensing by the KTN black hole. We first got the precession of the orbit for small declinations for the KTN black hole spacetime, which is very different from the Schwarzschild black hole case. Then we studied the lensing at small declination. The strong deflection limit coefficients were reconsidered and their behaviors were found to be different from the equatorial ones. Under this approximation, the positions and magnification of the images were studied. Furthermore, the critical curves and caustic structure were obtained. The critical points are close to the optical axis for the positive spin $a$ and farther for the negative spin $a$, and the nontrivial caustic structures acquire a finite extension. We also plotted the first six caustics for the KTN black hole for different spin $a$ and NUT charge $n$, respectively. Compared all these results to those for the Schwarzschild black hole and the Kerr black hole, it is easy to find that there is a significant effect of the NUT charge $n$ on the observables, i.e., the magnifying power $\bar{\mu}_{k}$, ratio $\bar{\mu}_{k+1}/\bar{\mu}_{k}$, and the intersections $\theta_{k}^{\textrm{cr}}$. Especially, for fixed value of spin $a$, the intersections $\theta_{k}^{\textrm{cr}}$ vary from $0.1\sim 2\mu$arcsec for different value of NUT charge. Such an optical resolution is reachable by very long baseline interferometry (VLBI) projects, and advanced radio interferometry between space and Earth (ARISE), which have the angular resolution of $10^{-3}$ arcsecond in the near infrared \cite{Eckart,BozzaMancini}. Thus, measuring $\theta_{k}^{\textrm{cr}}$ from astronomical observations, we are allowed to determine the value of the NUT charge $n$.

\section*{Acknowledgement}

This work was supported in part by the Program for New Century Excellent
Talents in University, the Huo Ying-Dong Education Foundation of
Chinese Ministry of Education (No. 121106), the National Natural
Science Foundation of China (No. 11075065), the National Natural Science Foundation of China (No. 11205074), and the Fundamental Research Funds for the Central Universities (No. lzujbky-2012-k30).

\end{document}